\documentclass[aps,prl,twocolumn]{revtex4}
\usepackage{amsmath}
\usepackage{amssymb}
\usepackage{natbib}
\usepackage{color}
\usepackage[dvips]{graphicx}

\begin{document}

\bibliographystyle{apsrev}

\title{An Asymmetric Elastic Rod Model for DNA}

\author{B. Eslami-Mossallam}
\affiliation{Department of Physics, Sharif University of
Technology, P.O. Box 11155-8639, Tehran,
Iran.}

\author{M.R. Ejtehadi}
\email{ejtehadi@sharif.edu}
\affiliation{Department of Physics, Sharif University of
Technology, P.O. Box 11155-8639, Tehran,
Iran.}

\date{\today}

\begin{abstract}
In this paper we consider the anharmonic corrections to the
anisotropic elastic rod model for DNA. Our model accounts for the
difference between the bending energies of positive and negative
rolls, which comes from the asymmetric structure of the DNA
molecule. We will show that the model can explain the high
flexibility of DNA at small length scales, as well as kink
formation at high deformation limit.
\end{abstract}


\maketitle

Characterizing the elastic behavior of DNA molecule is of crucial
importance in understanding its biological functions. In recent
years, single-molecule experiments such as DNA stretching and
cyclization \cite{Bustamante01, Crothers01} have provided us with
valuable information about the elasticity of long DNA molecules.
The results of these experiments can be described by the elastic
rod model (also called wormlike-chain model) \cite{Marko01,
Towles}. In this model it is assumed that the elastic energy is a
harmonic function of the deformation \cite{Marko03, Towles}. The
elastic rod model is very successful in explaining the elastic
behavior of the micron-size DNA molecules.

Recently, modern experimental techniques have made it possible to
study the elasticity of DNA at nanometer length scale
\cite{Widom01, Widom02, Wiggins01, Yuan}. In these experiments it
is observed that short DNA molecules are much more flexible than
predicted by the elastic rod model. Several different models have
been presented by now, that try to explain the origin of this
discrepancy by considering the possibility of local DNA melting
\cite{Marko02, Ranjith, Yuan, Palmeri}, or the occurrence of kinks
in the DNA structure \cite{Nelson01}. Also Wiggins \emph{et al.}
have suggested an alternative form for the elastic energy
\cite{Wiggins01, Wiggins02}.

Since the DNA is not a symmetric molecule, the energy required to
bend the DNA over its major groove is not equal to the energy
required to bend it over its minor groove. The model which in
introduced in this letter takes this difference into account. The
effect of asymmetric structure of DNA on its elastic energy has
been discussed previously by Marko and Siggia \cite{Marko03},
where they showed that there must be a coupling term between bend
and twist in the harmonic elastic energy. We will discuss that the
asymmetric structure of DNA can also be introduced as a
correction to the harmonic elastic energy, which is of the third
order. We shall show that our \emph{asymmetric elastic rod model}
can account for the high flexibility of short DNA molecules.

\begin{figure}[thb]
\begin{center}
\includegraphics[scale=0.5]{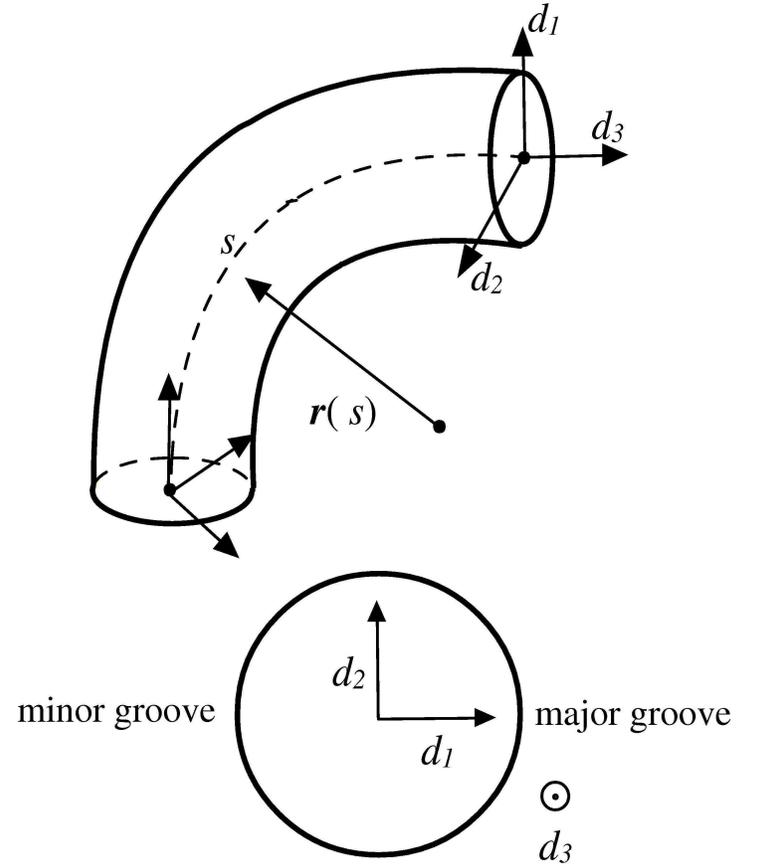}
\end{center} \caption{Parameterization of the elastic rod. The
local frame $\{\hat{d_{1}}, \hat{d_{2}}, \hat{d_{3}}\}$ is
attached to the rod. \label{fig:1}}
\end{figure}

In the elastic rod model DNA is represented by a continuous
inextensible rod. The curve which passes through the rod center
determines the configuration of the DNA in three dimensional
space. This curve is denoted by $\vec{r}$, and is parameterized
by the arc length parameter $s$ (Figure \ref{fig:1}). In
addition, a local coordinate system with an orthonormal basis
$\{\hat{d_{1}}, \hat{d_{2}}, \hat{d_{3}}\}$ is attached to each
point of the rod. As depicted in Figure \ref{fig:1},
$\hat{d_{3}}(s)$ is tangent to the curve $\vec{r}$ at each point,
$\hat{d_{3}}(s)=\mathrm{d}\vec{r}/\mathrm{d}s$, $\hat{d_{1}}(s)$
is perpendicular to $\hat{d_{3}}(s)$ and points toward the major
groove, and $\hat{d_{2}}(s)$ is defined as
$\hat{d_{2}}(s)=\hat{d_{3}}(s)\times \hat{d_{1}}(s)$.
These three orthogonal vectors uniquely determine the three
dimensional configuration of DNA. From classical mechanics we have
\begin{equation}
\label{ddot} \dot{\hat{d_i}}=\vec{\Omega}\times
\hat{d}_i\quad\quad\quad i=1,2,3\,,
\end{equation}
where the dot denotes the derivative with respect to $s$, and
$\vec{\Omega}$ is called the spatial angular velocity. The
components of $\vec{\Omega}$ in the local coordinate system are
denoted by $\Omega _{1}$, $\Omega_{2}$, and $\Omega_{3}$. The
elastic energy of an inextensible DNA in the most general form
can then be written as
\begin{equation}
\label{Energy-general} E=\int_0^L
\mathcal{E}[\Omega_1(s),\Omega_2(s),\Omega_3(s)]\,\mathrm{d}s\,,
\end{equation}
where $L$ is the total length of DNA. $\mathcal{E}(s)$ is the
energy per unit length of DNA, i.e the energy density, at point
$s$. For small deformations, the energy density can be written as
a Taylor expansion about the lowest energy configuration
\cite{Marko03}. For a DNA with no intrinsic curvature and a
constant intrinsic twist $\omega_0$, the lowest energy
configuration is given by $\vec{\Omega}_0=[0,0,\omega_0]^{\,T}$.
Thus, at the lowest order, we arrive at a harmonic energy density
in the form
\begin{equation}
\label{harmonicEnergy-general} \mathcal{E}_{\mathrm{harm}}
[\Omega_1,\Omega_2,\Omega_3]=\frac{1}{2}k_BT\,\vec{\omega}^{\,T}\,\mathbf{Q}\,\,\vec{\omega}\,,
\end{equation}
where $k_B$ is the Boltzmann constant, $T$ is the temperature,
and $\vec{\omega}$ is defined by
$\vec{\omega}=\vec{\Omega}-\vec{\Omega}_0$. $\mathbf{Q}$ is a
$3\times 3$ symmetric matrix whose elements are the elastic
constants of DNA \cite{Marko01, Towles, Becker}. Considering a
short segment of DNA with the length $\mathrm{d}s$ at the point
$s$, this segment has a symmetry under $180^{\circ}$ rotation
about the local $\hat{d_{1}}$ axis at the point $s$. Thus the odd
powers of $\Omega_1$ must not appear in the expansion of energy
density, and the matrix $\mathbf{Q}$ has only four independent
non-zero elements: $\mathbf{Q}_{11}$, $\mathbf{Q}_{22}$,
$\mathbf{Q}_{33}$, and $\mathbf{Q}_{23}$. Therefore, the harmonic
energy density can be written as \cite{Marko03}.
\begin{eqnarray}
\label{harmonicEnergy}\mathcal{E}_{\mathrm{harm}}=
\frac{1}{2}k_BT\big[A_1\,\Omega_1^2+A_2\,\Omega_2^2
+C(\Omega_3-\omega_0)^2\nonumber\\
+2D\,\Omega_2(\Omega_3-\omega_0)\big].\qquad\qquad
\end{eqnarray}
The first two terms in equation (\ref{harmonicEnergy}) correspond
to the bending of DNA over its grooves (roll), and over its
backbone (tilt), respectively. $A_1$ and $A_2$ are the
corresponding bending constants. Since roll  requires less energy
than tilt \cite{Calladine01, Lankas02, Behrouz}, one expects that
$A_2< A_1$. The third term indicates the energy needed for
twisting the DNA about its central axis, with the twist constant
$C$. Finally, the fourth term accounts for the coupling between
roll and twist \cite{Olson01}. Although the elastic constants of
DNA may depend on the sequence \cite{Becker}, in this paper we
neglect sequence dependence, and assume that they are constant
all along the DNA.

The existence of twist-roll coupling indicates that there is
indeed a difference between bending over major groove (positive
roll), and bending over minor groove (negative roll): For
positive values of $D$, the DNA has a tendency to untwist when
roll is positive, and to overtwist when roll is negative.\\
To account for the effect of asymmetry on the bending energy of
DNA, we need a term in the energy density which is an odd function
of $\Omega_2$, and does not depend on $\Omega_1$ or $\Omega_3$.
There is no such term in the harmonic elastic energy, so we
consider third-order terms in the expansion of energy density.
The term proportional to $\Omega_2^3$ has the desired property.
On the basis of some theoretical analysis \cite{Crick}, as well as
experimental evidences \cite{Richmond} and simulation studies
\cite{Lankas01}, we assume that negative roll is more favorable
than positive roll. Thus we write the third-order term in the form
$+1/3!\,F^2\,\Omega_2^3$, where $F$ is a real parameter. (It must
be noted that the main conclusion of the paper remains valid if
positive roll is easier than negative roll. To account for this
case, one can write the third-order term in the
form $-1/3!\,F^2\,\Omega_2^3$.)\\
To keep the model as simple as possible, we neglect couplings in
all orders, as well as higher-order corrections to the twist
energy. So the only third-order term which enters in the model is
$1/3!\,F^2\,\Omega_2^3$. Since the elastic energy must have a
lower bound, we must keep the fourth-order correction to the roll
energy, i. e. the term proportional to $\Omega_2^4$, in the
model. For consistency of the model, we also keep the
corresponding fourth-order correction to the tilt energy. Since
the anisotropy in bending energy is accounted for in the second
order, to reduce the model free parameters, we write the
fourth-order terms in the form
$1/4!\,G^3(\Omega_1^4+\Omega_2^4)$, with $G$ real and positive.
Adding third-order and fourth-order terms to the harmonic energy
density, we obtain the asymmetric elastic rod model which is
given by
\begin{eqnarray}
\label{asymmetricEnergy}\mathcal{E}_{\mathrm{asym}}=
\!k_BT\Big[\frac{1}{2}A_1\,\Omega_1^2+\frac{1}{2}A_2\,\Omega_2^2
+\frac{1}{2}C(\Omega_3-\omega_0)^2\nonumber\\
+\,\frac{1}{3\,!}\,F^{\,2}\,\Omega_2^3+\frac{1}{4\,!}\,G^{\,3}(\Omega_1^4+\Omega_2^4)\Big].\nonumber\\
\end{eqnarray}
This model accounts for the asymmetry between positive and
negative rolls, as well as the difference in the energies of roll
and tilt. Since there is no coupling term in the model, roll,
tilt, and twist can be regarded as independent deformations, and
the energy density can be decompose into three separate terms
\begin{equation}
\label{Energy-decomposed}
\mathcal{E}_{\mathrm{asym}}[\Omega_1,\Omega_2,\Omega_3]=\mathcal{E}_1[\Omega_1]+\mathcal{E}_2[\Omega_2]+\mathcal{E}_3[\Omega_3]\,,
\end{equation}
where
\begin{equation}
\label{E1} \mathcal{E}_{1}= k_BT\Big[\frac{1}{2}A_1\,\Omega_1^2+
\frac{1}{4\,!}\,G^{\,3}\,\Omega_1^4\Big],\qquad\qquad\qquad
\end{equation}
\begin{equation}
\label{E2} \mathcal{E}_{2}=
k_BT\Big[\frac{1}{2}A_2\,\Omega_2^2+\,\frac{1}{3\,!}\,F^{\,2}\,\Omega_2^3+
\frac{1}{4\,!}\,G^{\,3}\,\Omega_2^4\Big],\,\,
\end{equation}
\begin{equation}
\label{E3} \mathcal{E}_{3}= \frac{1}{2}k_BT\,
C(\Omega_3-\omega_0)^2\,.\qquad\qquad\qquad\qquad\quad
\end{equation}

\begin{figure}[thb]
\begin{center}
\includegraphics[scale=0.52]{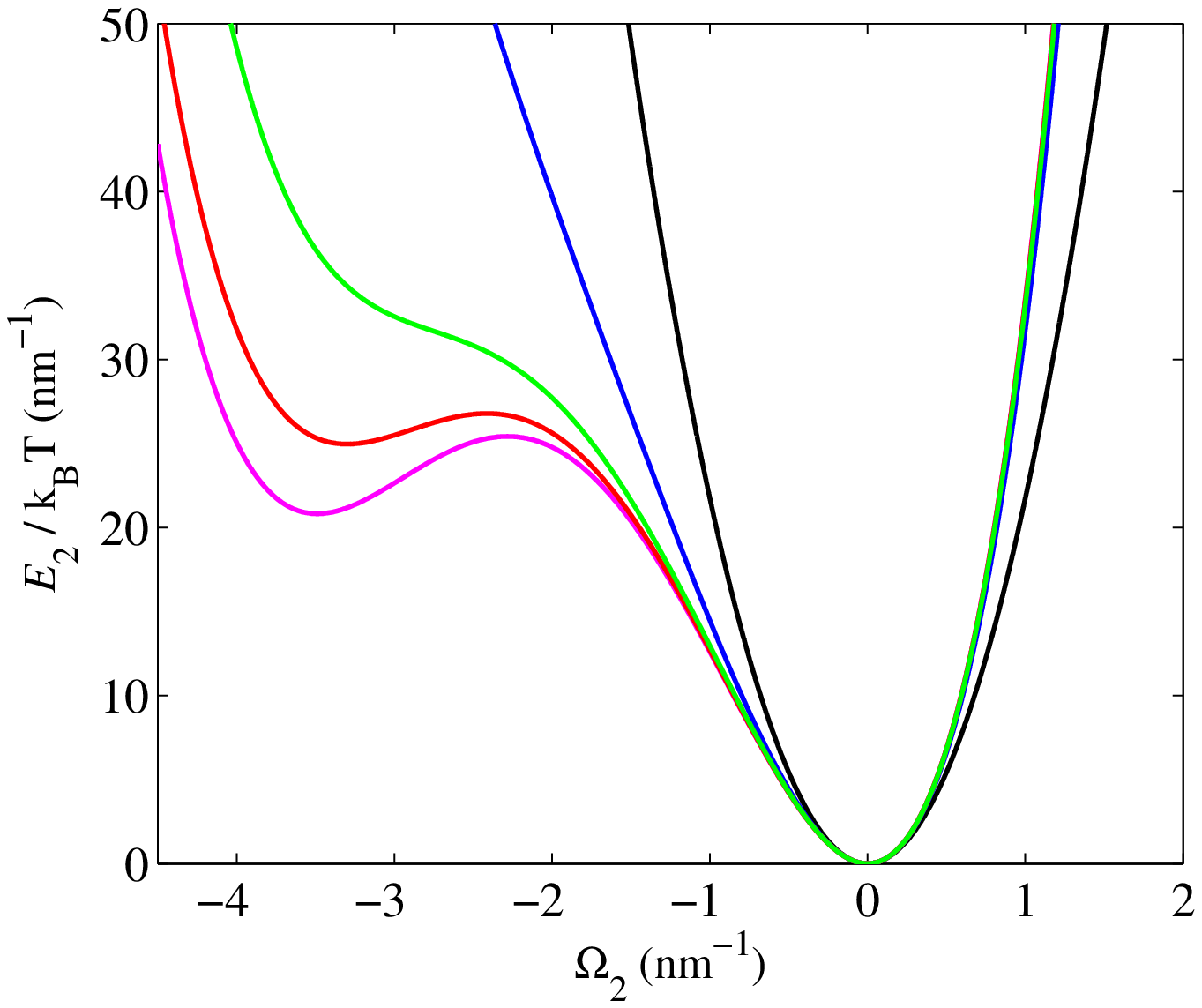}
\end{center}
\caption{(Color online) $\mathcal{E}_2$ as a function of
$\Omega_2$ for $A_2=43.50\,\mathrm{nm}$ and different values of
$F$ and $G$. Black: $F=G=0$. Blue: $G=3.20\,\mathrm{nm}$ and
$F=7.20\,\mathrm{nm}$, $\mathcal{E}_2$ has one minimum and its
curvature is positive every where. Green: $G=3.20\,\mathrm{nm}$
and $F=7.80\,\mathrm{nm}$, $\mathcal{E}_2$ has two inflection
points and one minimum. Red: $G=3.20\,\mathrm{nm}$ and
$F=7.90\,\mathrm{nm}$, $\mathcal{E}_2$ has two minima. Magenta:
$G=3.20\,\mathrm{nm}$ and $F=7.94\,\mathrm{nm}$. Note that a
change of $0.04 \,\mathrm{nm}$ in $F$ results in a change of
about $1\,k_BT/l_0$ in $\mathcal{E}_2$ in the vicinity of the
second minimum.
 \label{fig:2}}
\end{figure}

Depending on the values of $A_2$, $F$ and $G$, $\mathcal{E}_2$ can
take three different forms (see Figure \ref{fig:2}). For small
values of $F$, $\mathcal{E}_2$ has only one minimum at
$\Omega_2=0$ and its curvature is always positive. For $F>
(2\,A_2\,G^{\,3})^{1/4}$ the curvature of $\mathcal{E}_2$ can
change sign and there are two inflection points. For given $A_2$
and $G$ there exists an upper bound
$F_c=(8/3\,\,A_2\,G^{\,3})^{1/4}$, beyond which $\mathcal{E}_2$
has two minima: one at $\Omega_2=0$ and the other one at a
negative $\Omega_2$. In this case DNA has two stable
configurations, and there is always a barrier between them.
However, one expects that the barrier is not large compared to
$k_BT$ for a real DNA.

To study the elastic behavior of DNA in the asymmetric elastic
rod model, we calculate the distribution function $P(\theta)$,
the probability that the DNA bends into an angle $\theta $. We
use a Monte Carlo simulation to calculate $P(\theta)$. In this
simulation we discretize each chain into separate segments of
length $l_0=0.34\,\mathrm{nm}$, equal to the base-pair separation
in DNA. The orientation of each segment is then determined by a
vector $\vec{\Theta}$, where $|\vec{\Theta}|$ determines the
rotation angle of the local coordinate system with respect to the
laboratory coordinate system, and the direction of $\vec{\Theta}$
indicates the normal to the plane of rotation. The special
angular velocity is related to $\vec{\Theta}$ as
$\vec{\Omega}=\vec{\Theta}/l_0$. In each Monte Carlo move, we
randomly choose a segment along the chain, and for that segment
we change the vector $\vec{\Theta}$ by $\Delta\vec{\Theta}$. The
direction of $\Delta\vec{\Theta}$ is random, and its magnitude is
chosen randomly in the interval $[0\,,\,\Theta_0]$. $\Theta_0$ is
chosen so that the accept ratio is about $0.5$. We do not include
the self avoiding in the simulation, since the probability of self
crossing is small for the short simulated DNA molecules.

Recently, Wiggins \emph{et al.} have used atomic force microscopy
to measure distribution of the bending angle of short DNA
molecules \cite{Wiggins01}. Although the DNA molecules in Wiggins
\emph{et al.} experiment has the characteristic properties of two
dimensional polymers \cite{Wiggins01}, to simulate the experiment
we do not confine the DNA completely in a plane. The reason is
that the minimum energy configuration of an anisotropic DNA is
not plannar, although the deviation from a plannar configuration
is negligible \cite{Maddocks01}. It is known in the anisotropic
harmonic elastic rod model, that the effective bending constant
of long DNA molecules in three dimensions is equal to the
harmonic average of $A_1$ and $A_2$ \cite{Behrouz, Becker,
Lankas02}, $
A_{\mathrm{eff}}=\left[1/2\left(1/A_1+1/A_2\right)\right]^{-1}$,
while in the two dimensions the effective bending constant is
given by $A_{\mathrm{eff}}^{2D}=\sqrt{A_1\,A_2}$ \cite{Golnoosh}.
Since $A_{\mathrm{eff}}^{2D}$ is always greater than
$A_{\mathrm{eff}}$, confining the DNA in a plane costs energy.
For this reason, we allow the DNA to come out of the plane by
$0.3\,\mathrm{nm}$, which is seven times smaller than DNA
diameter and lies in the range of atomic length scales.

\begin{figure}[thb]
\begin{center}
\includegraphics[scale=0.52]{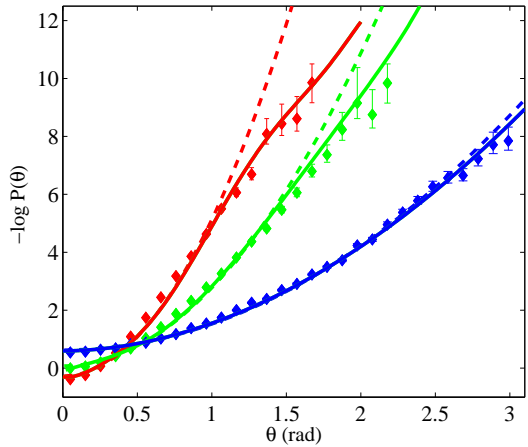}
\end{center}
\caption{(Color online) $-\log P(\theta)$ as a function of
$\theta$, for $L=5\,\mathrm{nm}$ (red), $L=10\,\mathrm{nm}$
(green), and $L=30\,\mathrm{nm}$ (blue). Dots show the
experimental data of Wiggins \emph{et al.} \cite{Wiggins01}.
Curves show the theoretical results. Dashed curves: isotropic
harmonic elastic rod model, with $A_1=A_2=54\,\mathrm{nm}$, and
$F=G=0$. Solid curves: $A_1=87.00\,\mathrm{nm}$,
$A_2=43.50\,\mathrm{nm}$ $F=7.90\,\mathrm{nm}$, and
$G=3.20\,\mathrm{nm}$.\label{fig:3}}
\end{figure}

Following other studies \cite{Towles, Nelson02}, we assume
$A_1=2\,A_2$, $C=100\,\mathrm{nm}$, and
$\omega_0=1.8\,\mathrm{nm^{-1}}$. The values of $A_2$, $F$ and
$G$ are then determined by fitting the theoretical results to the
experimental data of Wiggins \emph{et al.}, with the constrain
that the persistence length of the DNA is $54\,\mathrm{nm}$
\cite{Wiggins01}. Figure \ref{fig:3} shows a good fit to the
experimental data, which corresponds to $A_2=43.50\,\mathrm{nm}$
$G=3.20\,\mathrm{nm}$ $F=7.90\,\mathrm{nm}$. The predictions of
isotropic-harmonic elastic rod model are also shown in the
Figure. We report the values of the fitting parameter with three
significant digits. The reason is that $\mathcal{E}_2$ is very
sensitive to the changes of the parameters, specially when it has
two minima. In fact, a change in the order of
$10^{-2}\,\mathrm{nm}$ in these parameters may results in a
change in  $\mathcal{E}_2$ in the order of $1\,k_BT/\l_0$ (see
Figure \ref{fig:2}), and therefore can significantly affect the
elastic behavior of DNA. We must note here that the ratio
$A_1/A_2$ is also relevant to the fitting procedure. However, one
can obtain equally good fits for different values of $A_1/A_2$.

The functional form of $\mathcal{E}_2$ for
$A_2=43.50\,\mathrm{nm}$, $F=7.90\,\mathrm{nm}$ and
$G=3.20\,\mathrm{nm}$ is shown in Figure \ref{fig:2}, which has
two minima. The second minimum occurs at
$\Omega_2=-3.3\,\mathrm{nm}^{-1}$ which corresponds to a
$-64^{\circ}$ roll between adjacent base pairs. Thus, the
existence of a second minimum can lead to the formation of kinks
in the minor groove direction in a tightly bent DNA.
 As can be seen in Figure \ref{fig:3},
both in the experiment and our model, $-\log P(\theta)$ deviates
from the harmonic model at large bending angles. Continuing the
graphs in our model, they arrive to an approximately linear
regime. This linear behavior is a signature of kink formation.
Both the slope of the line,
and the crossover point are related to the values of $F$ and $G$.\\
The possibility of kink formation in the DNA structure has been
considered previously by other authors. It was firstly mentioned
by Crick and Klug, who proposed an atomistic structure for a
kinked DNA \cite{Crick}. They suggest that DNA can be kink most
easily toward the minor groove. Nelson, Wiggins, and Phillips
have presented a simple model for kinkable elastic rods
\cite{Nelson01}, in which the kinks are completely flexible, and
can be formed in any direction with equal probability. Their
model can explain the high cyclization probability of short DNA
molecules \cite{Widom01, Widom02}. The linear behavior is also
observed in their model \cite{Nelson01}, but the slope of the line
is always zero. In a recent experiment, Du \emph{et al.} proved
the existence of kinks in DNA minicircles of 64-65 bp
\cite{Vologodskii01}. Molecular dynamics simulations on a 94 bp
minicircle \cite{Lankas01} also show that kinks are formed, with
the same structure predicted by Crick and Klug. Similar kinks in
the direction of minor groove have been observed in the
structure of nucleosomal DNA \cite{Richmond}.\\
Kinks are also observed in the crystal structures of non-histone
protein-DNA complexes \cite{Dickerson, Olson01}. In these
complexes, DNA has a clear tendency to kink in the major groove
direction. Du \emph{et al.} have found the distribution function
$P(\theta)$ for a base-pair step in these complexes
\cite{Vologodskii02}. Although it is contradictory to our primary
assumption, that kinks are formed toward the major groove, the
asymmetric elastic rod model can be fitted to the Du \emph{et
al.} data by writing the third-order term in the form
$-1/3!\,F^2\,\Omega_2^3$, and choosing $F=8.86\,\mathrm{nm}$ and
$G=3.83\,\mathrm{nm}$. We found that the model with these values
of $F$ and $G$ can not explain the experimental data of Wiggins
\emph{et al.}. This shows that the statistical property of DNA in
the protein-DNA complexes certainly differs from the free DNA.
This difference is probably due to the interaction of proteins
with DNA, which can alter the DNA conformation dramatically, and
leads to different effective elastic constants.

In this paper, we presented a generalization of the anisotropic
elastic rod model, assuming that the energies of positive and
negative rolls are different as a result of the asymmetric
structure of DNA. We showed that this model can explain the
elastic behavior of short DNA molecules. We also showed that this
model allows the formation of kinks in the DNA structure when the
DNA is tightly bent. The kinks always form in one of the groove
directions, as suggested by other studies.
\begin{acknowledgments}
We are highly indebted to P. Nelson for providing us with the
experimental data presented in Fig. 3. We also thank R.
Golestanian for encouraging comments, R. Everaers and N. Becker
for valuable discussions, H. Amirkhani for her comments on the
draft manuscript, and the Center of Excellence in Complex Systems
and Condensed Matter (CSCM) for partial support.
\end{acknowledgments}
\bibliography{biblio}
\end{document}